\documentclass[aps,pra,twocolumn,superscriptaddress,longbibliography]{revtex4-1}
\usepackage{etex}
\usepackage{amsmath,amssymb,amsthm}
\usepackage[colorlinks=true,citecolor=blue,urlcolor=blue]{hyperref}
\usepackage[pdftex]{graphicx}
\usepackage{times,txfonts}
\usepackage{braket}
\usepackage{natbib}
\usepackage{amssymb,amsmath,bm,amsbsy,amscd}
\usepackage{mathtools}
\usepackage{latexsym}
\usepackage{subfig}
\usepackage{tabularx, booktabs}
\usepackage{graphics,epstopdf}
\usepackage{graphicx,color}
\usepackage{float}
\usepackage{graphicx}
\usepackage{amsfonts}
\usepackage{color,soul}

\begin{document}

%\preprint{}

% user-defined command 
\newcommand{\bv}{\mathbf{v}}
\newcommand{\bh}{\mathbf{h}}
\newcommand{\bz}{\mathbf{z}}

%------------------------------------------------------------------------------------------------------------%
\title{Entropy, Free Energy, and Work of Restricted Boltzmann Machines}
\author{Sangchul Oh}
\email[]{soh@hbku.edu.qa}
\affiliation{Qatar Environment and Energy Research Institute, 
Hamad Bin Khalifa University, Qatar Foundation, P.O.Box 5825, Doha, Qatar}
\author{Abdelkader Baggag}
\email[]{abaggag@hbku.edu.qa}
\affiliation{Qatar Computing Research Institute, 
Hamad Bin Khalifa University, Qatar Foundation, P.O.Box 5825, Doha, Qatar}
\author{Hyunchul Nha}
\email[]{hyunchul.nha@qatar.tamu.edu}
\affiliation{Department of Physics, Texas A\&M University at Qatar, 
Education City, P.O.Box 23874, Doha, Qatar}
%\thanks{}

\date{\today}

%------------------------------------------------------------------------------------------------------------%
\begin{abstract}
A restricted Boltzmann machine is a generative probabilistic graphic network. A probability of finding the 
network in a certain configuration is given by the Boltzmann distribution. Given training data, its learning 
is done by optimizing parameters of the energy function of the network. In this paper, we analyze 
the training process of the restricted Boltzmann machine in the context of statistical physics. As an 
illustration, for small size Bar-and-Stripe patterns, we calculate thermodynamic quantities such as entropy, 
free energy, and internal energy as a function of training epoch. We demonstrate the growth of the correlation 
between the visible and hidden layers via the subadditivity of entropies as the training proceeds. Using 
the Monte-Carlo simulation of trajectories of the visible and hidden vectors in configuration space, we also 
calculate the distribution of the work done on the restricted Boltzmann machine by switching the parameters 
of the energy function. We discuss the Jarzynski equality which connects the path average of the exponential
function of the work and the difference in free energies before and after training. 
\end{abstract}

%\showkeys
\maketitle

%------------------------------------------------------------------------------------------------------------%
\section{Introduction}
A restricted Boltzmann machine (RBM)~\cite{Smolensky1986} is a generative probabilistic neural network. 
RBMs and general Boltzmann machines are described by a probability distribution with parameters, 
i.e., the Boltzmann distribution. An RBM is an undirected Markov random fields, and 
is considered a basic building block of deep neural network. RBMs have been applied widely, for example, 
dimensional reduction, classification, feature learning, pattern recognition, topic modeling, and so 
on~\cite{Hinton2012,Fischer2014,Melchior2016}. 

%Our studies may provide a more complete picture of learning process together with a novel insight. 
%It has wide applications from the
%mage generation to the neural network representation of quantum many-body states.

As its name implies, a RBM is closely connected to physics, and they shares some important concepts such 
as entropy, free energy, etc.~\cite{Mehta2019}. Recently, RBMs have renewed much attention in physics since 
Carleo and Troyer~\cite{Carleo2017} showed that a quantum many-body state could be efficiently represented 
by the RBM.  
Gabr\'e {\it et al.} and Tramel {\it et al.}~\cite{Tramel2018} employed the 
Thouless-Anderson-Palmer mean-field approximation, used for a spin glass problem, to replace the Gibbs 
sampling of contrast-divergence training. Amin {\it et al.}~\cite{Amin2018} proposed a quantum Boltzmann 
machine based on quantum Boltzmann distribution of a quantum Hamiltonian. More interestingly, there is a deep 
connection between Boltzmann machine and tensor networks of quantum many-body 
systems~\cite{Stoudenmire2016,Gao2017,Chen2018,Sarma2019,Huggins2019}.
Xia and Kais combined the restricted Boltzmann machine and quantum algorithms to calculate
the electronic energy of small molecules~\cite{Xia2018}.

While the working principles of RBMs have been well established, it may be still needed to understand the RBM 
better for further applications. In this paper, we investigate the RBM from the perspective of statistical 
physics. As an illustration, for bar-and-stripe pattern data, the thermodynamic quantities such as the entropy,
the internal energy, the free energy, and the work, are calculated as a function of epoch. Since the RBM is a bipartite 
system composed of visible and hidden layers, it may be interesting, and informative, to see how the correlation 
between the two layers grows as the training goes on. We show that the total entropy of the RBM is always less 
than the sum of the entropies of visible and hidden layers, except at the initial time when the training begins. 
This is the so-called subadditivity of entropy, indicating that the visible layer becomes correlated with 
the hidden layer as the training proceeds. The training of the RBM is to adjust the parameters of the energy 
function, which can be considered as the work done on the RBM, from a thermodynamic point of view. 
Using the Monte-Carlo simulation of the trajectories of the visible and hidden vectors in configuration space,
we calculate the work of a single trajectory and the statistics of the work over the ensemble of trajectories.
We also examine the Jarzynski equality that connects the ensemble of the work done on the RBM  
and the difference in free energies before and after training of the RBM.

The paper is organized as follows. In Section~\ref{Sec:RBM}, the detailed analysis of the RBM from 
the statistical physics point of view is described. In Section~\ref{Sec:summary}, we presents 
the summary of the result together with discussions.

%----------------------------------------------------------------------------------------------------------------------%
\section{Statistical Physics of Restricted Boltzmann Machines}
\label{Sec:RBM}

\subsection{Restricted Boltzmann machines}
% Figure 1
\begin{figure}[h]
\centering{\includegraphics[width=0.5\textwidth,angle=0]{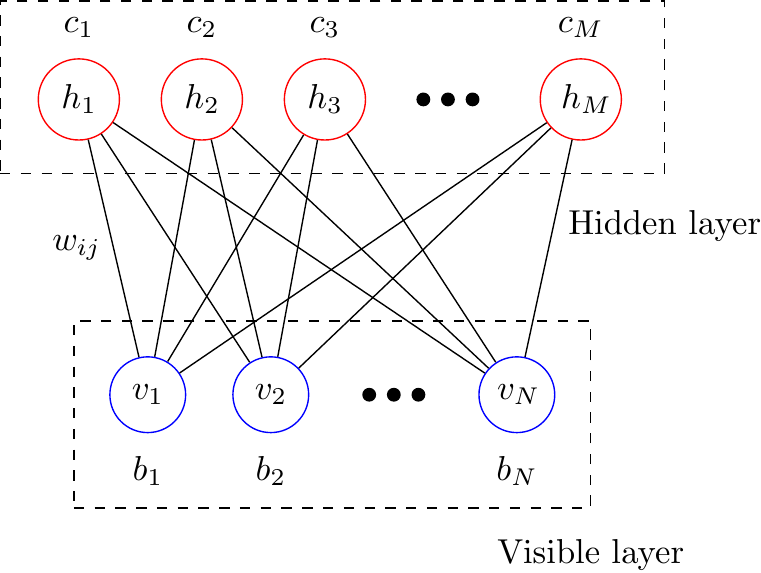}}
\caption{Graph structure of a restricted Boltzmann machine with the visible layer and the hidden layer.}
\label{Fig1}
\end{figure}

Let us start with a brief introduction of the RBM~\cite{Smolensky1986,Hinton2012,Fischer2014}.
As shown in Fig.~(\ref{Fig1}), the RBM is composed of two layers; the visible layer and the hidden layer.
Possible configurations of the visible and hidden layers are represented by the random binary vectors, 
$\bv=(v_1,\dots,v_N) \in \{0,1\}^N$ and $\bh = (h_1,\dots,h_M) \in \{0,1\}^M$, respectively. 
The interaction between the visible and hidden layers is given by the so-called weight matrix $w\in
\mathbb{R}^{N}\times\mathbb{R}^M$ where the weight $w_{ij}$ is the connection strength between 
a visible unit $v_i$ and a hidden unit $h_j$. The biases $b_i\in\mathbb{R}.$ and $c_j\in\mathbb{R}$ are 
applied to visible unit $i$ and hidden unit $j$, respectively. Given random vectors $\bv$ and $\bh$, 
the energy function of the RBM is written as an Ising-type Hamiltonian
\begin{equation}
E(\bv,\bh;\theta) = -\sum_{i=1}^N\sum_{j=1}^M w_{ij}v_ih_j -\sum_{i=1}^{N} b_iv_i -\sum_{i=1}^{M}c_i h_i\,,
\end{equation}
where the set of model parameters is denoted by $\theta\equiv\{w_{ij}, b_i, c_j\}$.
The joint probability of finding $\bv$ and $\bh$ of the RBM is given by the Boltzmann distribution
\begin{equation}
p(\bv,\bh;\theta) = \frac{e^{-E(\bv,\bh;\theta)}}{Z},
\label{Eq:Boltzmann_distribution}
\end{equation}
where the partition function, $Z(\theta) \equiv \sum_{\bv,\bh} e^{-E(\bv,\bh;\theta)}$, is the sum over 
all possible configurations.  The marginal probabilities $p(\bv;\theta)$ and $p(\bh;\theta)$ for visible 
and hidden layers are obtained by summing up the hidden or visible variables, respectively,
\begin{subequations}
\begin{align}
p(\bv;\theta) 
&= \sum_{\bh} p(\bv,\bh;\theta) = \frac{1}{Z(\theta)} \sum_{\bh} e^{-E(\bv,\bh;\theta)}\,,\\
p(\bh;\theta) 
&= \sum_{\bv} p(\bv,\bh;\theta) = \frac{1}{Z(\theta)} \sum_{\bv} e^{-E(\bv,\bh;\theta)}\,.
\end{align}
\end{subequations}

The training of the RBM is to adjust the model parameter $\theta$ such that the marginal probability of the 
visible layer $p(\bv;\theta)$ becomes as close as possible to the unknown probability $p_{\rm data}(\bv)$ 
that generate the training data. Given identically and independently sampled training data 
${\cal D} \in \{\bv^{(1)},\dots,\bv^{(D)}\}$, the optimal model parameters $\theta$ can be obtained by 
maximizing the likelihood function of the parameters, 
${\cal L}(\theta|{\cal D}) = \prod_{i=1}^D p(\bv^{(i)};\theta)$, or equivalently by maximizing the 
log-likelihood function $\ln{\cal L} (\theta|{\cal D}) = \sum_{i=1}^D \ln p(\bv^{(i)};\theta)$. 
Maximizing the likelihood function is equivalent to minimizing the Kullback-Leibler divergence or the 
relative entropy of $p(\bv;\theta)$ from $q(\bv)$~\cite{Kullback1951,Cover2006} 
\begin{equation}
D_{\rm KL}(q\,||\,p) = \sum_{\bv} q(\bv)\ln\frac{q(\bv)}{p(\bv;\theta)}\,,\\
\label{Eq:KL_divergence}
\end{equation}
where $q(\bv)$ is an unknown probability that generates the training data. 
Another method of monitoring the progress of training is the cross-entropy cost between 
the input visible vector $\bv^{(i)}$ and a reconstructed visible vector $\bar{\bv}^{(i)}$ of the RBM,
\begin{equation}
C = -\frac{1}{D}\sum_{i\in D}\left[
    \bv^{(i)} \ln \bar{\bv}^{(i)} + (1-\bv^{(i)}) \ln(1-\bar{\bv}^{(i)})
    \right]\,.
\label{Eq:cross_entropy}
\end{equation}
The stochastic gradient ascent method for the log-likelihood function is used to train the RBM. Estimating
the log-likelihood function requires the Monte-Carlo sampling for the model probability distribution.
Well-known sampling methods are the contrast-divergence, denoted by CD-$k$, and the persistent contrast 
divergence PCD-$k$. For the detail of the RBM algorithm, please see Refs.~\cite{Hinton2012,Fischer2014,Melchior2016}.
Here we employ the CD-$k$ method.

%------------------------------------------------------------------------------------------------------------%
\subsection{Free energy, entropy, and internal energy}
From physics point of view, the RBM is a finite classical system composed of two subsystems, similar to an 
Ising spin system. The training of the RBM is considered the driving of the system from an initial equilibrium 
state to the target equilibrium state by switching the model parameters. It may be interesting to see how 
thermodynamic quantities such as free energy, entropy, internal energy, and work, change as the training progresses.

It is straightforward to write down various thermodynamic quantities for the total system. The free energy $F$ 
is given by the logarithm of the partition function $Z$,
\begin{equation}
F(\theta) = -\ln Z(\theta)\,.
\label{Eq:free_energy}
\end{equation}
The internal energy $U$ is given by the expectation value of the energy function $E(\bv,\bh;\theta)$
\begin{equation}
U(\theta) = \sum_{\bv,\bh} E(\bv,\bh;\theta) p(\bv,\bh;\theta)\,.
\label{Eq:internal_energy}
\end{equation}
The entropy $S$ of the total system comprising the hidden and visible layers is given by
\begin{equation}
S(\theta) = -\sum_{\bv,\bh} p(\bv,\bh;\theta)\ln p(\bv,\bh;\theta) \,.
\label{Eq:total_entropy}
\end{equation}
Here, the convention of $0\ln 0 = 1 $ is employed if $p(\bv,\bh) = 0$. The free energy ~(\ref{Eq:free_energy})
is related to the difference between the internal energy~(\ref{Eq:total_entropy}) and the 
entropy~(\ref{Eq:FUTS}) 
\begin{equation}
F = U -TS\,,
\label{Eq:FUTS}
\end{equation}
where $T$ is set to 1. 

\begin{figure}[ht]
\includegraphics[width=0.45\textwidth]{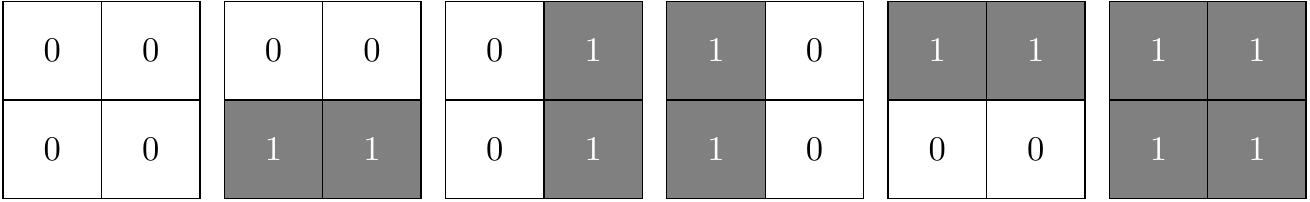}
\caption{6 samples of $2\times2$ Bar-and-Stripe patterns used as the training data in this work. 
Each configuration is represented by a visible vector $\bv \in \{0,1\}^{2\times2}$ 
or by a decimal number; $(0,0,0,0) = 0$, $(0,0,1,1) = 3$, $(0,1,0,1)=5$, $(1,0,1,0) =10$, $(1,1,0,0)=12$, 
$(1,1,1,1) =15$ in row-major ordering.} 
\label{Fig:Bar_stripe}
\end{figure}

Generally, it is very challenging to calculate the thermodynamic quantities, even numerically. The number of possible 
configurations of $N$ visible units and $M$ hidden units grow exponentially as $2^{N+M}$. Here, for a feasible 
benchmark test, the $2\times2$ Bar-and-Stripe data are considered ~\cite{Hinton1986,MacKay2002}. 
Fig.~\ref{Fig:Bar_stripe} shows the 6 possible $2\times2$ Bar-and-Stripe patterns out of 16 possible configurations,
which will be used as the training data in this work. We take the sizes of the visible and the hidden layers as 
$N=4$ and $M=6$, respectively. In order to understand better how the RBM is trained, the thermodynamic quantities 
are calculated numerically for this small benchmark system.

\begin{figure}[ht]
\includegraphics[width=0.5\textwidth]{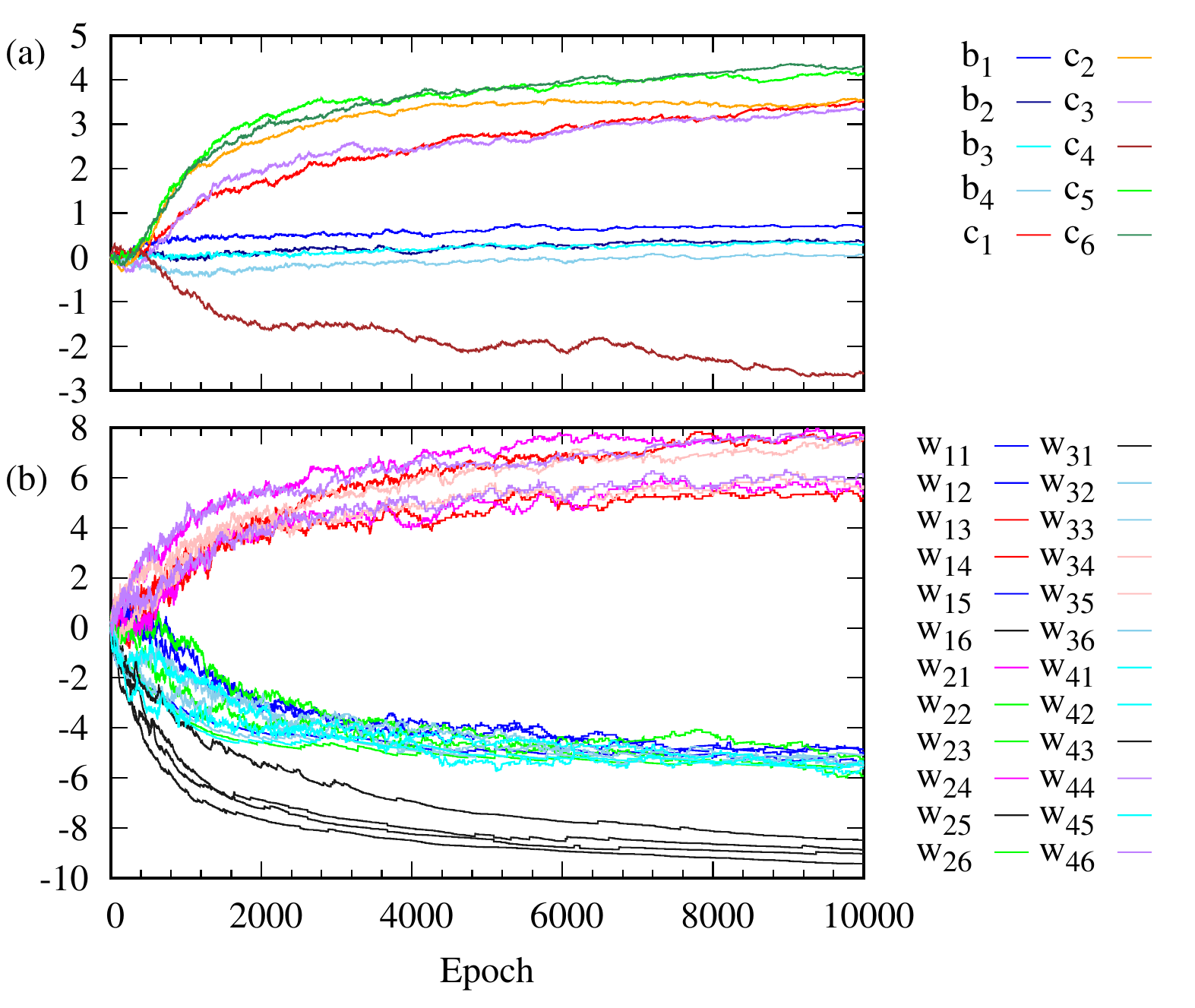}
\caption{(a) Bias $b_i$ on the visible unit $i$ and bias $c_j$ on the hidden unit $j$ are plotted as a function 
of epoch. (b) Weight $w_{ij}$ connecting the visible unit $i$ and the hidden unit $j$ are plotted as 
a function of epoch.}
\label{parameters_epoch}
\end{figure}
\begin{figure}[ht]
\includegraphics[width=0.4\textwidth]{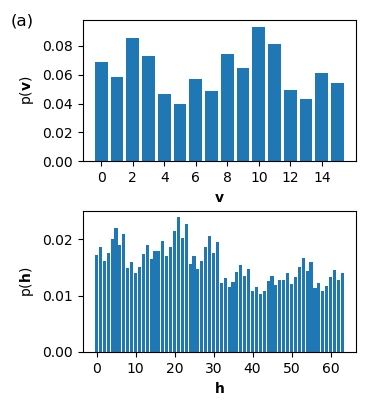}
\includegraphics[width=0.4\textwidth]{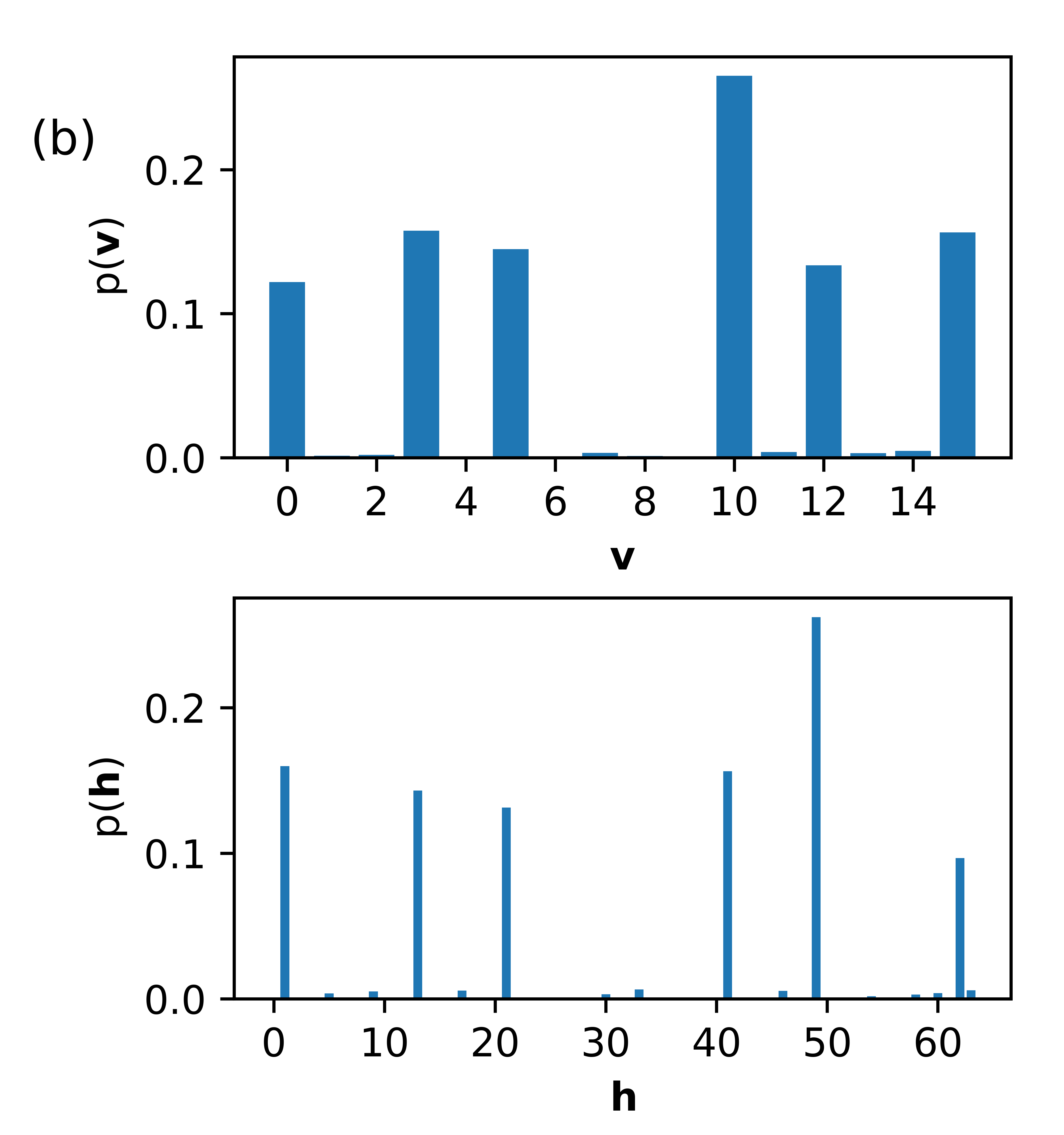}
\caption{Marginal probabilities $p(\bv)$ of visible layer and $p(\bh)$ of hidden layer are plotted
(a) before training and (b) after training. 
The binary vector $\bv$ or $\bh$ in x-axis  is represented by the decimal number 
as noted in the caption of Fig.~\ref{Fig:Bar_stripe}. 
The visible and the hidden layers have total number of configurations given by $2^4=16$ and $2^6=64$, 
respectively. The learning rate is 0.15, the training epoch 20000, and CD-$k$ 100.} 
\label{Fig:probability}
\end{figure}

Fig.~\ref{parameters_epoch} shows how the weight $w_{ij}$, the bias $b_i$ on the visible unit $i$ 
and the bias $c_j$ on the hidden unit $j$ change as the training goes on. The weight $w_{ij}$ are 
clustered into 3 classes. The evolution of the bias $b_i$ on the visible layer is somewhat different from
that of the bias $c_j$ on the hidden layer. The change in $c_i$ are larger than that in $b_i$.
Fig.~\ref{Fig:probability} shows the change in the marginal 
probabilities $p(\bv)$ of the visible layer and $p(\bh)$ of the hidden layer before and after training. Note 
that the marginal probability $p(\bv)$ after training is not distributed exclusively over 6 possible outcomes
according to the training data set in Fig.~\ref{Fig:Bar_stripe}.

\begin{figure}[ht]
\includegraphics[width=0.5\textwidth]{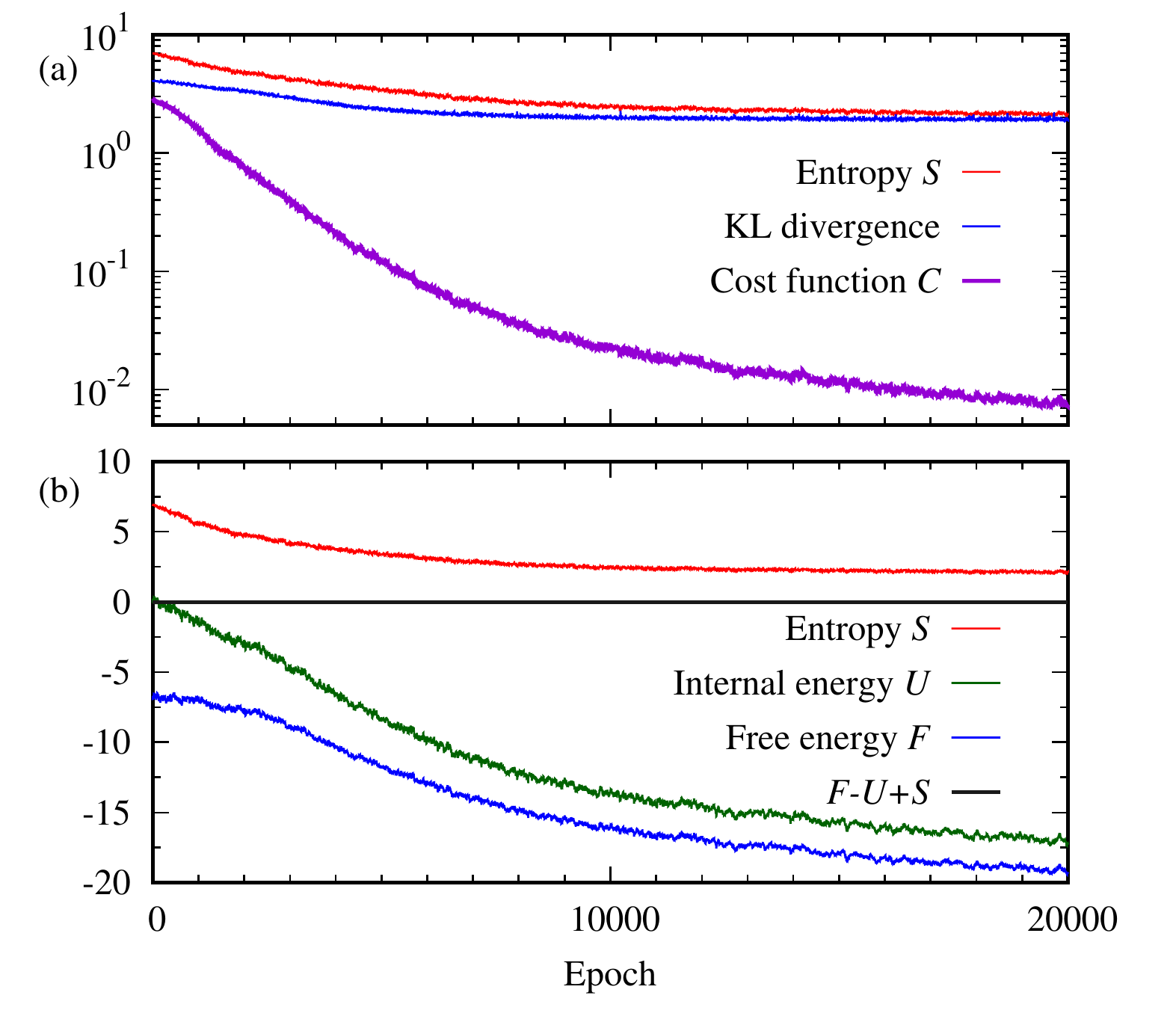}
\caption{For $2\times2$ bar-and-stripe data, (a) cost function $C$, entropy $S$, and the Kullback-Leibler 
divergence $D_{\rm KL}(q\,||\,p)$ are plotted as a function of epoch. (b) Free energy $F$, entropy $S$, and internal 
energy $U$ of the RBM are calculated as a function of epoch.}
\label{Fig:Thermodynamic_functions}
\end{figure}

Typically, the progress of learning of the RBM is monitored by the loss function. Here the Kullback-Leibler 
divergence, Eq.~(\ref{Eq:KL_divergence}) and the reconstructed cross entropy, Eq.~(\ref{Eq:cross_entropy}) 
are are used. Fig.~\ref{Fig:Thermodynamic_functions} plots the reconstructed cross
entropy $C$, the Kullback-Leibler divergence $D_{\rm KL}$, the entropy $S$, the free energy $F$, 
and the internal energy $U$ as a function of epoch. As shown in Fig.~\ref{Fig:Thermodynamic_functions} (a),
it is interesting to see that even after a large number of epochs $\sim10,000$, the cost function $C$ continues 
approaching zero while the entropy $S$ and the Kullback-Leibler divergence $D_{\rm KL}$ become steady.
On the other hand, the free energy $F$ continues decreasing together with the internal energy $U$, as 
depicted in Fig.~\ref{Fig:Thermodynamic_functions} (b). The Kullback-Leibler divergence is a well-known 
indicator to the performance of RBMs. Then, our result implies that the entropy may be another 
good indicator to monitor the progress of RBM while other thermodynamic quantities may be not.

In addition to the thermodynamic quantities of the total system of the RBM, Eqs.~(\ref{Eq:free_energy}), 
(\ref{Eq:total_entropy}), and (\ref{Eq:internal_energy}), it is interesting to see how the two subsystems of 
the RBM evolve. Since the RBM has no intra-layer connection, the correlation between the visible layer and the 
hidden layer may increase as the training proceeds. The correlation between the visible layer and the hidden 
layer can be measured by the difference between the total entropy and the sum of the entropies of the two 
subsystems. The entropies of the visible and hidden layers are given by 
\begin{subequations}
\begin{align}
S_V &=-\sum_{\bv} p(\bv;\theta) \ln p(\bv;\theta)\,,\\ 
S_H &=-\sum_{\bh} p(\bh;\theta) \ln p(\bh;\theta)\,.
\end{align}
\end{subequations}

The entropy $S_V$ of the visible layer is closely related to the Kullback-Leibler divergence of $p(\bv;\theta)$ 
to an unknown probability $q(\bv)$ which produces the data. Eq.~(\ref{Eq:KL_divergence}) is expanded as
\begin{align}
D_{\rm KL}(q\,||\,p) 
= \sum_{\bv} q(\bv)\ln q(\bv) - \sum_{\bv} q(\bv)\ln p(\bv;\theta)\,.
\end{align}
The second term $-\sum_{\bv} q(\bv)\ln p(\bv;\theta)$ depends on the parameter $\theta$. As the training proceeds,
$p(\bv;\theta)$ becomes close to $q(\bv)$ so the behavior of the second term is very similar to that of the entropy
$S_V$ of the visible layer. If the training is perfect, we have $q(\bv)=p(\bv;\theta)$ that leads to $D_{\rm KL}(q\,||\,p)=0$ while $S_V$ remains nonzero.

\begin{figure}[ht]
\includegraphics[width=0.5\textwidth]{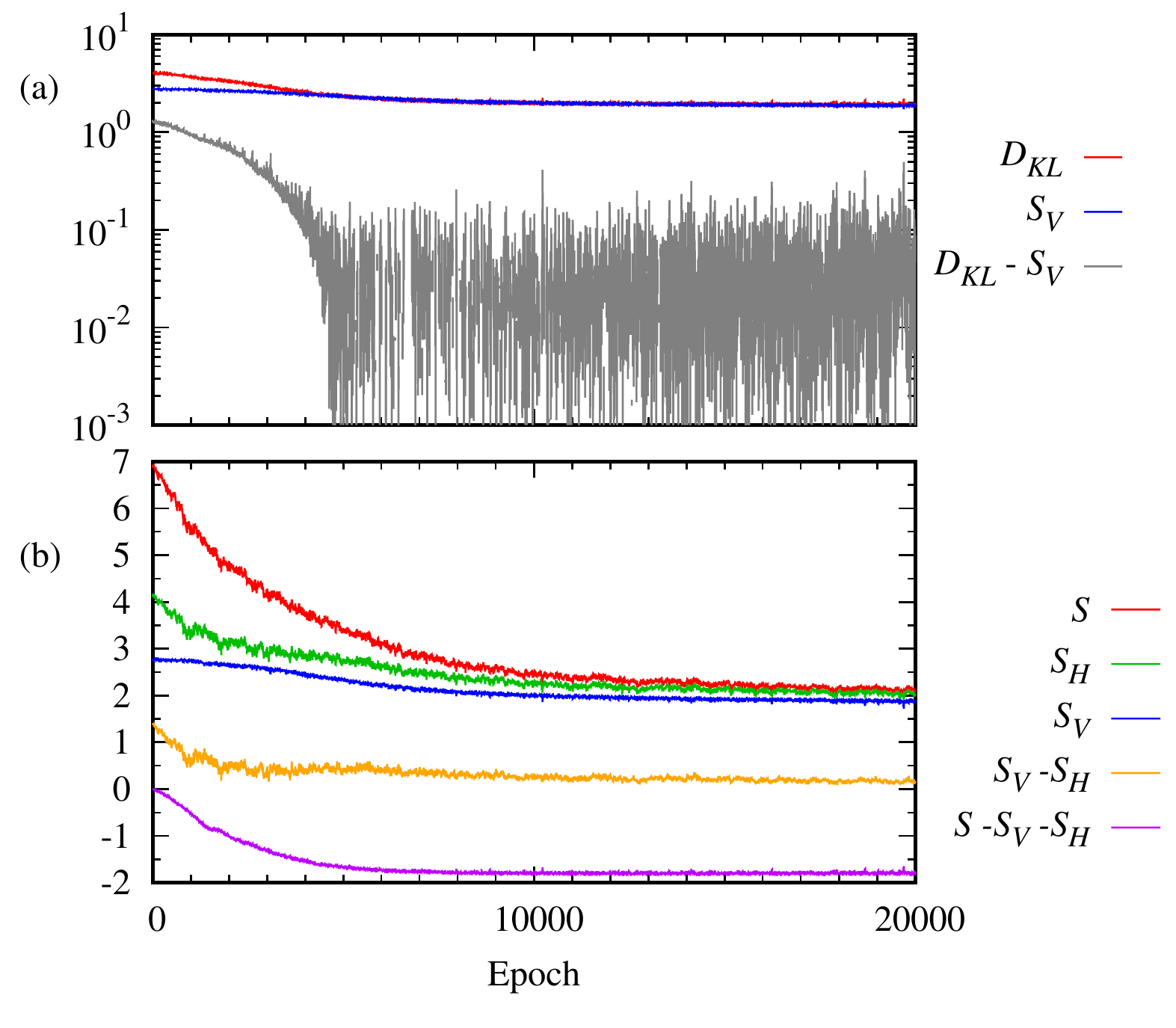}
\caption{(a) Kullback-Leibler divergence $D_{\rm KL}(q\,||\,p)$, entropy $S_V$, and their difference are 
plotted as a function of epoch. (b) Entropy $S$ of the total system, entropy $S_V$ of the visible layer, 
entropy $S_H$ of the hidden layer, and the difference $S-S_H-S_V$ are plotted as a function of epoch.}
\label{Fig:Entropies}
\end{figure}

The difference between the total entropy and the sum of the entropies of subsystems is written as
\begin{equation}
S -(S_V + S_H) = \sum_{\bv,\bh} p(\bv,\bh) \ln\left[ \frac{p(\bv)\,p(\bh)}{p(\bv,\bh)} \right]\,.
\label{Eq:Entropy_sum}
\end{equation}
Eq.~(\ref{Eq:Entropy_sum}) tells that if the visible random vector $\bv$ and the hidden random vector
$\bh$ are independent, i.e., $p(\bv,\bh;\theta) = p(\bv;\theta)p(\bh;\theta)$, then the entropy $S$ of 
the total system is the sum of the entropies of subsystems. 
In general, the entropy $S$ of the total system is always less than or equal to the sum of the entropy of 
the visible layer, $S_V$, and the entropy of the hidden layer, $S_H$,~\cite{Reif1965},
\begin{equation}
S \le S_V + S_H\,.
\label{Eq:Entropy_subadditivity}
\end{equation}
This is called the subadditivity of entropy, one of the basic properties of the Shannon entropy, which 
is also valid for the von Neumann entropy~\cite{Araki1970,Nielsen2010}.
This property can be proved using the log inequality, 
$-\ln x\ge -x +1$. In other way, Eq.~(\ref{Eq:Entropy_subadditivity}) may be proved by using the log-sum 
inequality, which states that for the two sets of nonnegative numbers, $a_1,\dots,a_n$ and $b_1,\dots,b_n$,
\begin{equation}
\sum_ia_i\log\frac{a_i}{b_i} \ge \left(\sum_i a_i\right) \log\frac{\left(\sum_i a_i\right)}{\left(\sum_i b_i\right)}
\end{equation}
In other words, Eq.~(\ref{Eq:Entropy_sum}) can be regarded as the negative of the relative entropy or 
Kullback-Leibler divergence of the joint probability $p(\bv,\bh)$ to 
the product probability $p(\bv)\cdot p(\bh)$,
\begin{equation}
I\bigl( p(\bv,\bh)\,||\,p(\bv)p(\bh) \bigr)
= \sum_{\bv,\bh} p(\bv,\bh)\log\left[\frac{p(\bv,\bh)}{p(\bv)p(\bh)} \right]\,.
\end{equation}

For the 2$\times 2$ Bar-and-Stripe pattern, the entropies of visible and hidden layers, $S_V, S_H$ are 
calculated numerically. Fig.~\ref{Fig:Entropies} plots the entropies, $S_V, S_H$, $S$, and the Kullback-Leibler
divergence $D_{\rm KL}(q\,||\,p)$ as a function of epoch. Fig.~\ref{Fig:Entropies} (a) shows that 
the Kullback-Leibler divergence, $D_{\rm KL}(q\,||\,p)$ becomes saturated, though above zero, 
as the training proceeds.  Similarly, the entropy $S_V$ of the visible layer is saturated. This implies 
that the entropy of the visible layer, as well as the total entropy shown in 
Fig.~\ref{Fig:Thermodynamic_functions}, can be an indicator to learning better than the reconstructed 
cross entropy $C$, Eq.~(\ref{Eq:cross_entropy}). The same can also be said about the entropy of the hidden 
layer, $S_H$.  

The difference between the total entropy and the sum of the entropies of the two subsystems, $S-(S_V+S_H)$, 
becomes less than $0$, as shown in Fig.~\ref{Fig:Entropies} (b). Thus it demonstrates the subadditivity of 
entropy, i.e., the correlation between the visible and the hidden layer as the training proceeds. 
As it is saturated just as the total entropy and the entropies of the visible and hidden layers after 
a large number of epoch, the correlation between the visible layer and the hidden layer can also be a good 
quantifier of the RBM progress.

%----------------------------------------------------------------------------------------------------------------------%
\subsection{Work, free energy, and Jarzynski equality}

The training of the RBM may be viewed as driving a finite classical spin system from an initial equilibrium 
state to a final equilibrium state by changing the system parameters $\theta$ slowly. If the parameters 
$\theta$ are switched infinitely slowly, the classical system remains in quasi-static equilibrium.
In this case, the total work done on the systems is equal to the Helmholtz free energy difference between 
the before-training and the after-training, $W_\infty = F_1 - F_0 \,.$
For switching $\theta$ at a finite rate, the system may not evolve immediately to an equilibrium state,
the work done on the system depends on a specific path of the system in the configuration space. 
%The ensemble average of the work over the paths is greater than or equal to the free energy difference 
%$\langle W\rangle_\text{path} \ge \Delta F$ between the initial and final equilibrium states.
Jarzynski~\cite{Jarzynski1997,Jarzynski2011} proved that for any switching rate the free energy difference 
$\Delta F$ is related to the average of the exponential function of the amount of work $W$ over the paths
\begin{equation}
\langle e^{-W}\rangle_{\rm path} = e^{-\Delta F} \,.
\label{Eq:Jarzynski}
\end{equation}

The RBM is trained by changing the parameters $\theta$ through a sequence 
$\{\theta_0,\theta_1,\dots,\theta_{\tau}\}$, as shown in Fig.~\ref{parameters_epoch}. To calculate 
the work done during the training, we perform the Monte Carlo simulation of the trajectory of 
a state $(\bv,\bh)$ of the RBM in configuration space. From the initial configuration, 
$(\bv_0,\bh_0)$ which is sampled from the initial Boltzmann distribution, 
Eq.~(\ref{Eq:Boltzmann_distribution}), the trajectory 
$(\bv_0,\bh_0) \to (\bv_1,\bh_1)\to \cdots \to (\bv_{\tau},\bh_{\tau})$ is obtained using 
the Metropolis-Hastings algorithm of the Markov chain Monte-Carlo method~\cite{Metropolis1953,Hastings1970}. 
Assuming the evolution is 
Markovian, the probability of taking a specific trajectory is the product of the transition 
probabilities at each step, 
\begin{align}
p(\bv_0,\bh_0 \stackrel{\theta_1}{\longrightarrow} \bv_1,\bh_1)\,
p(\bv_1,\bh_1 \stackrel{\theta_2}{\longrightarrow} \bv_1,\bh_1)& \nonumber\\
\dots p(\bv_{\tau-1},\bh_{\tau-1}\stackrel{\theta_\tau}{\longrightarrow}\bv_{\tau},\bh_{\tau})&\,.
\end{align}
The transition $(\bv,\bh) \to (\bv',\bh')$ can be implemented by the Metropolis-Hastings algorithm 
based on the detailed balance condition for the fixed parameter $\theta$,
\begin{align}
\frac{ p(\bv,\bh \stackrel{\theta}{\longrightarrow} \bv',\bh') }
     { p(\bv,\bh \stackrel{\theta}{\longleftarrow} \bv',\bh')  }
= \frac{e^{-E(\bv',\bh';\theta)}}{e^{-E(\bv,\bh;\theta)}} \,.
\end{align}
The work done on the RBM at epoch $i$ may be given by
\begin{equation}
\delta W_i = E(\bv_i,\bh_i;\theta_{i+1}) - E(\bv_i,\bh_i;\theta_i)\,.
\end{equation}
The total work $W=\sum\delta W_i$ performed on the system is written as~\cite{Crooks1998} 
\begin{align}
W &= \sum_{i=0}^{\tau-1}\left[ E(\bv_i,\bh_i;\theta_{i+1}) -E(\bv_i,\bh_i;\theta_i)\right] \,.
\label{Eq:total_work}
\end{align}

\begin{figure}[ht]
\includegraphics[width=0.5\textwidth]{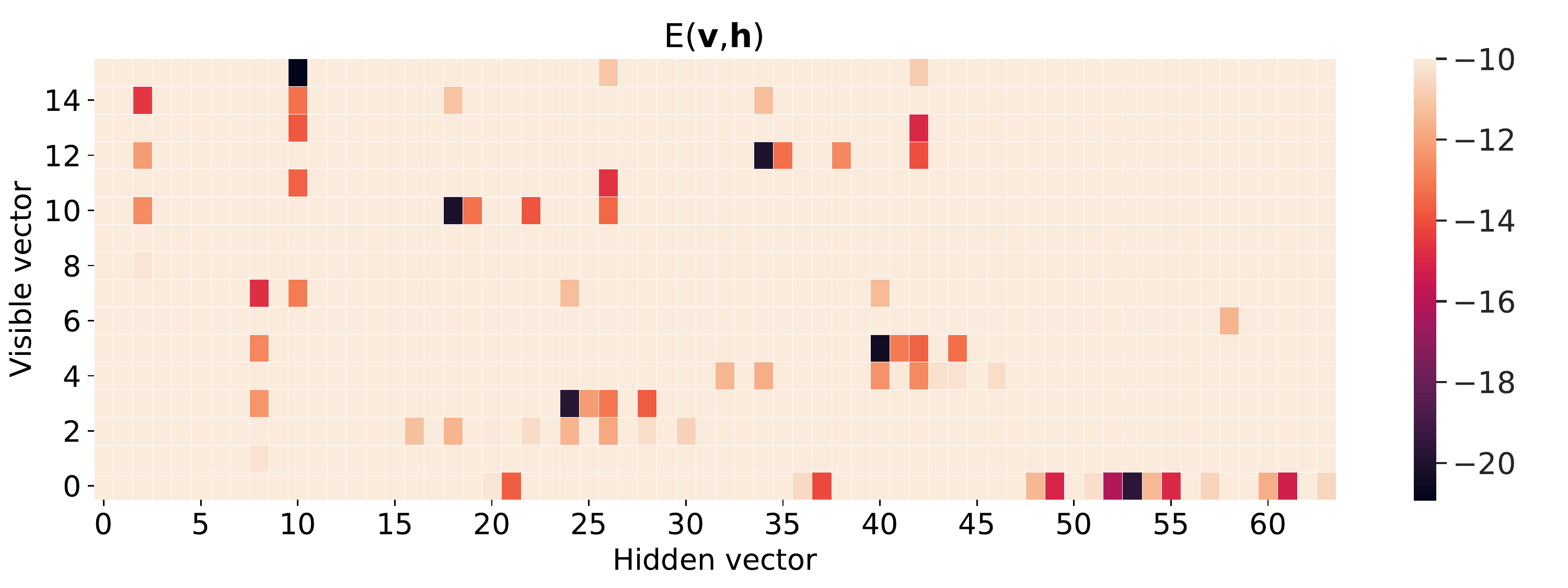}
\caption{Heat map of energy function $E(\bv,\bh;\theta)$, representing the energy level of each configuration,
after training of $2\times2$ Bar-and-Stripe patterns for 50000 epochs. The sizes of visible and hidden layers are
$N=4$ and $M=6$, respectively. The learning rate is $r=0.15$ and the value of CD$_k$ is $k=100$. The vertical and 
the horizontal axes represent each configuration of the visible and the hidden layers, respectively. 
The black tiles represent the lowest energy configurations among all configurations, thus 
the probability of finding that configuration is high.}
\label{Fig:Energy_levels}
\end{figure}

\begin{figure}[ht]
\includegraphics[width=0.5\textwidth]{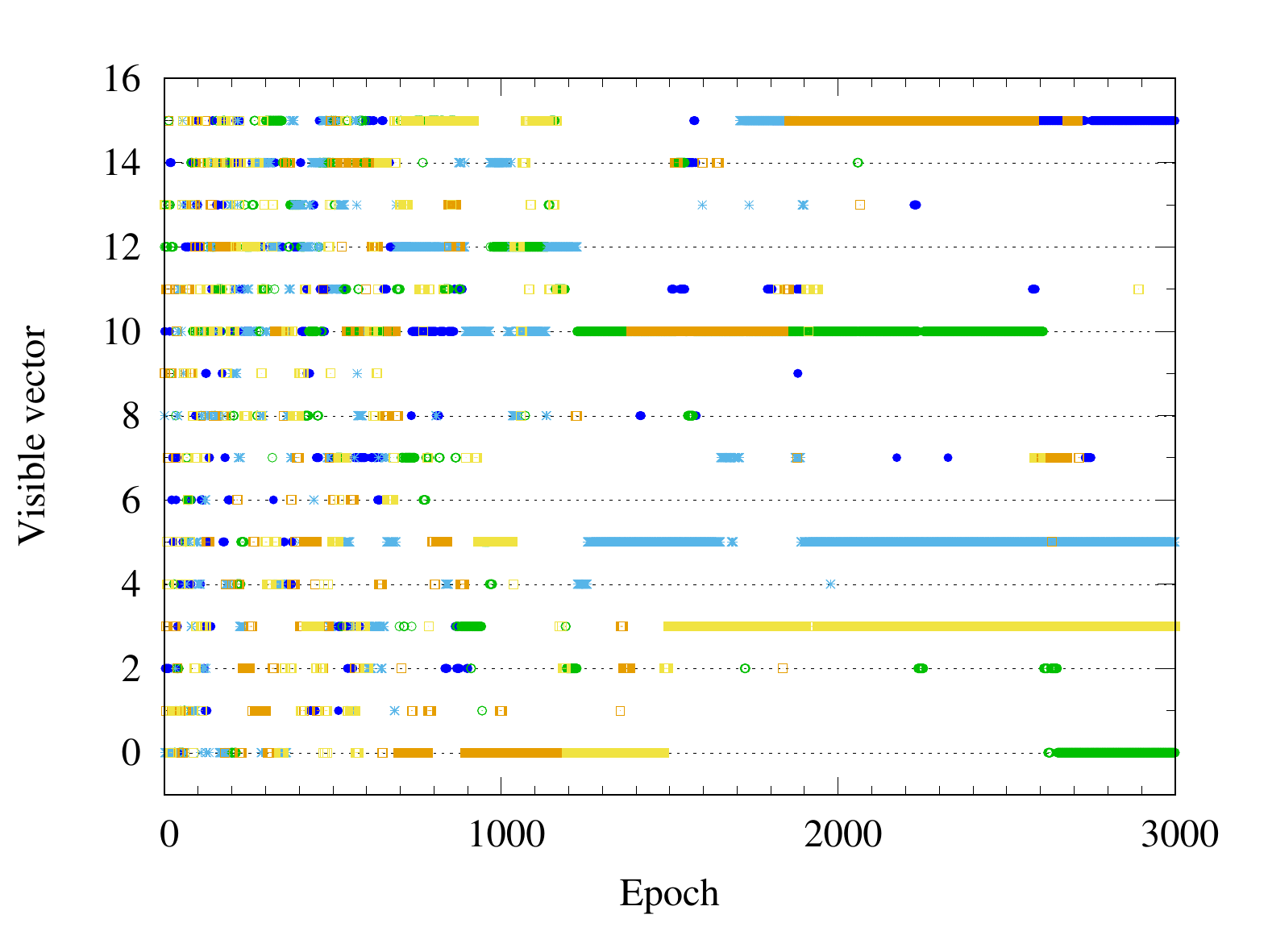}
\caption{Markov chain Monte-Carlo trajectories of the visible vector $\bv_i$ are plotted as a function 
of epoch.  The visible vector jumps frequently in the early state of training and becomes trapped into one 
of target states as the training proceeds.}
\label{Fig:Trajectory}
\end{figure}

Given the sequence of the model parameter $\{\theta_0,\theta_1,\dots,\theta_\tau\}$, the Markov evolution of the 
visible and hidden vectors $(\bv,\bh)\in \{0,1\}^{N+M}$ may be considered the discrete random walk. Random walkers
move to the points with low energy in configuration space.  Fig.~\ref{Fig:Energy_levels} shows the heat map of 
energy function $E(\bv,\bh;\theta)$ of the RBM for the $2\times 2$ Bar-and-Stripe pattern after training. One can 
see the energy function has deep levels at the visible vectors corresponding to the Bar-and-Stripe patterns of
the training data set in Fig.~\ref{Fig:Bar_stripe}, representing probability of generating the trained patterns 
is high. Also note that the energy function has many local minima. Fig.~\ref{Fig:Trajectory} plots 
a few Monte-Carlo trajectories of the visible vector $\bv$ as a function of epoch. Before training, 
the visible vector $\bv$ is distributed over all possible configurations, represented by the number 
$(0,\cdots, 15)$.  As the training progresses, the visible vector $\bv$ becomes trapped
into one of the six possible outcomes $(0,3,5,10,12,15)$. 

\begin{figure}[ht]
\includegraphics[width=0.5\textwidth]{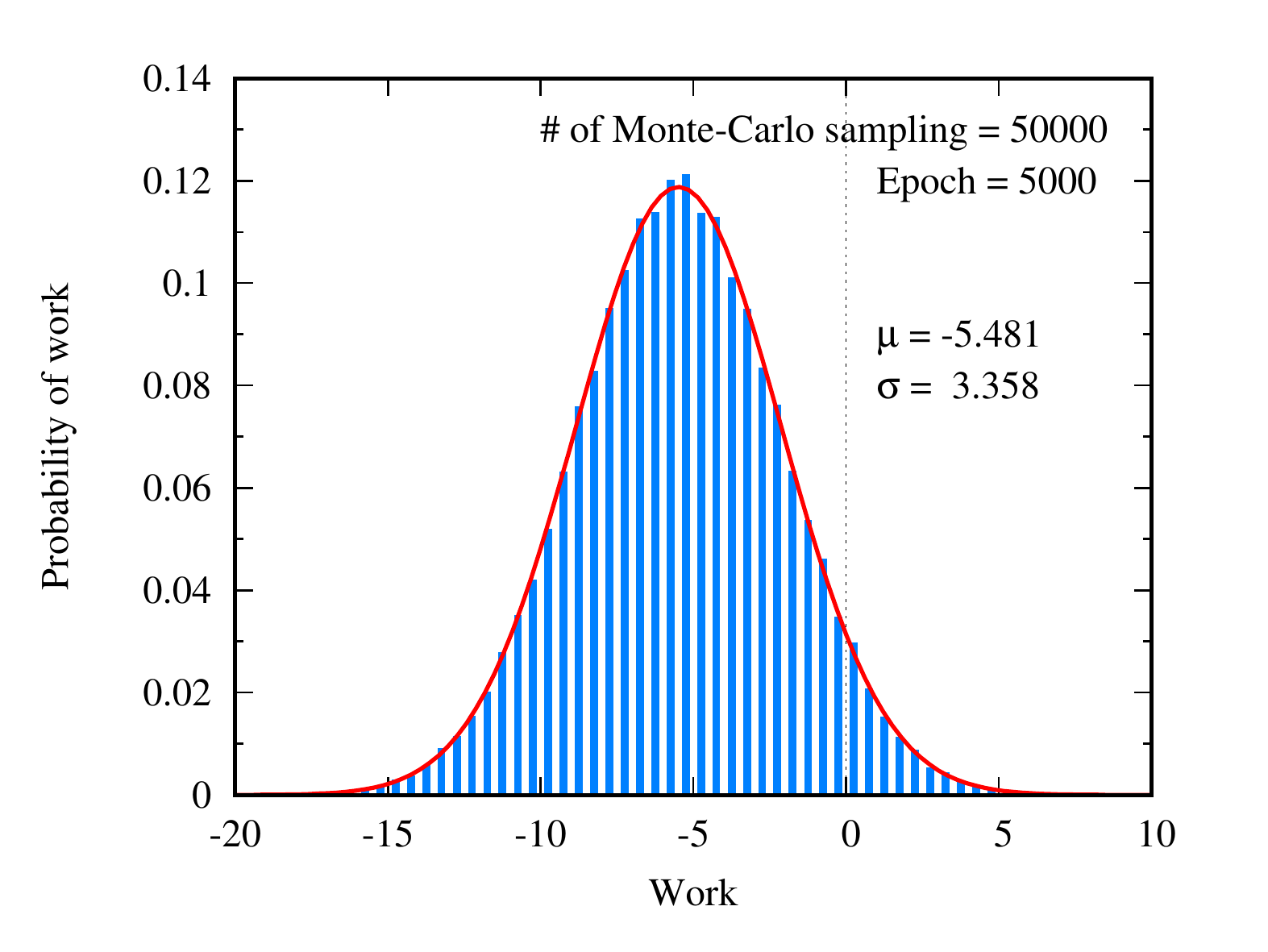}
\caption{Gaussian distribution of work done by the RBM during the training. The number of the Monte-Carlo sampling 
is 50000. The red curve is the plot of the Gaussian distribution using
the mean and the standard deviation calculated by the Monte-Carlo simulation.}
\label{Fig:work}
\end{figure}

\begin{figure}[ht]
\includegraphics[width=0.5\textwidth]{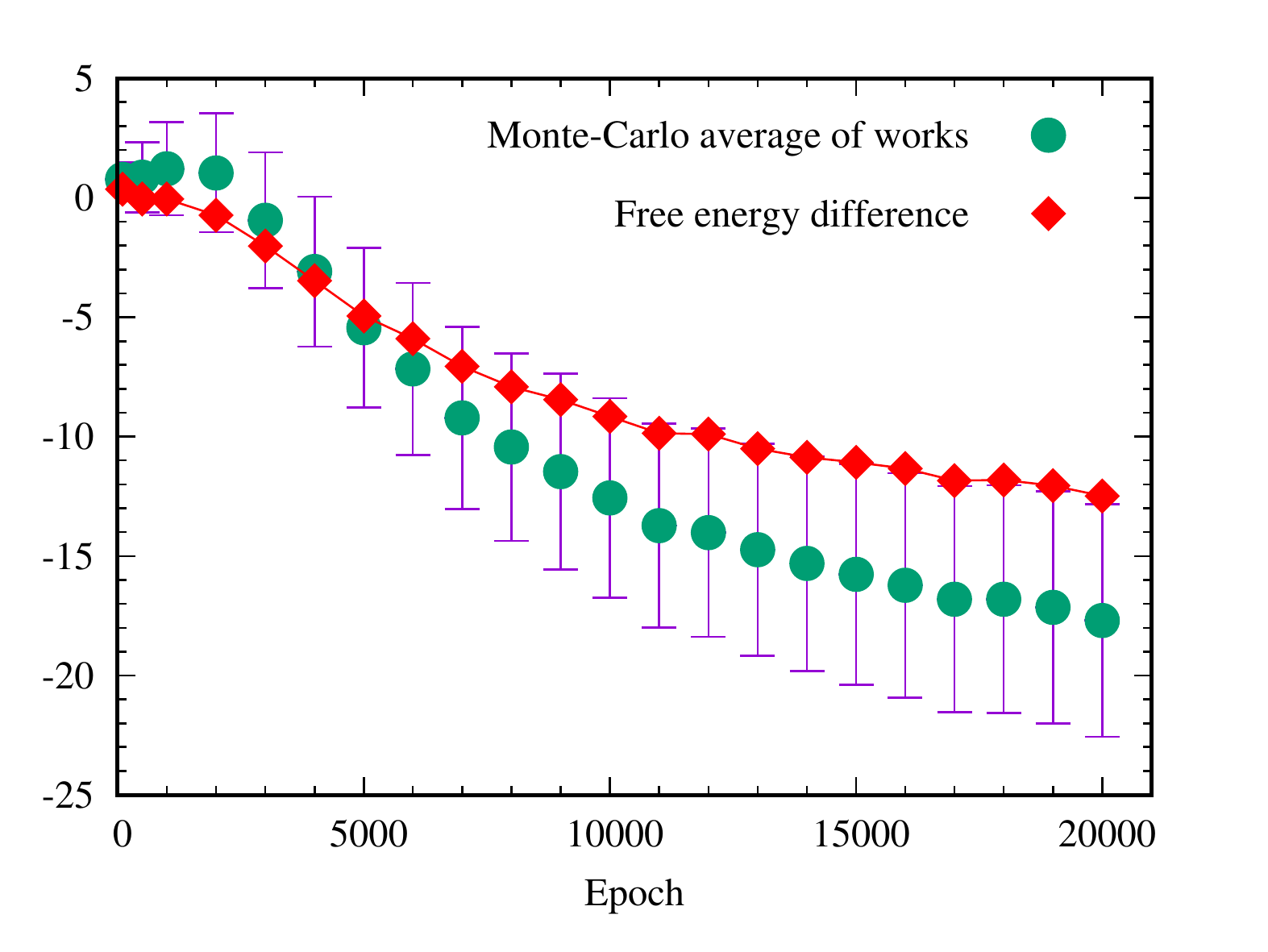}
\caption{Average of work done  with standard deviation and free energy difference 
$\Delta F = F({\rm epoch}) -F({\rm epoch}=0)$ as a function of epoch. 
The error bar of the work represent the standard deviation
of the Gaussian distribution. 
}
\label{Fig:work_free}
\end{figure}

In order to examine the relation between work done on the RBM during the training and 
the free energy difference, the Monte-Carlo simulation is performed to calculate the average of 
the work over paths generated by the Metropolis-Hastings algorithm of the Markov chain Monte-Carlo method. 
Each path starts from an initial state sampled from the uniform distribution over configuration space, 
as shown in Fig.~\ref{Fig:probability} (a). Since the work done on the system
depends on the path, the distribution of the work is calculated by generating many trajectories.
Fig.~\ref{Fig:work} shows the distribution of the work over 50000 paths at 5000 training epoch.
The Monte-Carlo average of the work is $\langle{W}\rangle \approx -5.481$, and its standard deviation 
is $\sigma_W \approx 3.358$. The distribution of the work generated by the Monte-Carlo simulation is well 
fitted to the Gaussian distribution, as depicted by the red curve in Fig.~\ref{Fig:work}.
This agrees with the statement of in Ref.~\cite{Jarzynski2011} that for the slow switching of the model 
parameters the probability distribution of work is approximated to the Gaussian. 

We perform the Monte-Carlo calculation of the exponential average of work, $\langle{e^{-W}} \rangle_{\rm path}$ 
to check the Jarzynski equality, Eq.~(\ref{Eq:Jarzynski}). The free energy difference can be estimated 
as 
\begin{equation}
e^{-\Delta F} =\langle e^{-W}\rangle_{\rm path} \approx \frac{1}{N_{\rm mc}}\sum_{n=1}^{N_{\rm mc}} e^{-W_n} \,,
\end{equation}
where $N_{\rm mc}$ is the number of the Monte-Carlo samplings. 
At small epoch number, the Monte-Carlo estimated value of free energy difference is close to $\Delta F$ 
calculated from the partition function. However, this Monte-Carlo calculation gives rise to the poor estimation 
of the free energy difference if the epoch is greater than 5000. This numerical errors can be explained by the 
fact that the exponential average of the work is dominated by rare 
realization~\cite{Jarzynski2006,Zuckerman2002,Lechner2006,Lechner2007,Halperin2016}. As shown in 
Fig.~\ref{Fig:work}, the distribution of work is given by the Gaussian distribution $\rho(W)$ with the mean 
$\langle W \rangle$ and the standard deviation $\sigma_W$. If the standard deviation $\sigma_W$ becomes larger, 
the peak position of $\rho(W) e^{-W}$ moves to the long tail of the Gaussian distribution. So the main contrition 
of the integration of $\langle e^{-W}\rangle$ comes from the rare realizations. Fig.~\ref{Fig:work_free} shows 
that the standard deviation $\sigma_W$ grows with epoch, so the error of the Monte-Carlo estimation of the
exponential average of the work grows quickly.

If $\sigma_W^2 \ll k_BT$, the free energy is related to the average of work and its variance as 
\begin{equation}
\Delta F = \langle{W}\rangle_{\rm path} -\frac{\sigma_W^2}{2k_BT} \,.
\label{}
\end{equation}
\smallskip
Here, the case is opposite, the spread of the value of work is large, i.e., $\sigma_W^2 \gg k_BT$ $(=1)$, 
so the central limit theorem does not work and the above equation can not be applied~\cite{Hendrix2001}.
Fig.~\ref{Fig:work_free} shows how the average of work, $\langle W \rangle_{\rm path}$, over 
the Markov chain Monte-Carlo paths changes as a function of epoch. The standard deviation of the Gaussian 
distribution of the work also grows as a function of training epoch. The free energy difference between 
before-training and after-training is called the reversible work $W_r = \Delta F$. The difference between 
the actual work and the reversible work is called the dissipative work, $W_d = W -W_r$~\cite{Crooks1998}. 
As depicted in Fig.~\ref{Fig:work_free},
the magnitude of the dissipative work grows with training epoch.

%----------------------------------------------------------------------------------------------------------------------%
\section{Summary}
\label{Sec:summary}
In summary, we analyzed the training process of the RBM in the context of statistical physics. 
In addition to the typical loss function, i.e., the reconstructed cross entropy, the thermodynamic 
quantities such as free energy $F$, internal energy $U$, and entropy $S$ were calculated as a function 
of epoch. While the free energy  and the internal energy decrease rather indefinitely with epochs, 
the total entropy and the entropies of the visible and the hidden layers become saturated together 
with the Kullback-Leibler divergence after a sufficient number of epochs. This result suggests that the entropy of 
the system may be a good indicator to the RBM progress along with Kullback-Leibler divergence.
It seems worth investigating the entropy for other larger data sets, for example, MNIST handwritten 
digits~\cite{MNIST2010}, in future works.

We have further demonstrated the subadditivity of the entropy, i.e., the entropy of the total 
system is less than the sum of the entropies of the two layers. This manifested the correlation 
between the visible and hidden layers growing with the training progress. Just as the entropies are 
well saturated together with Kullback-Leibler divergence, so is the correlation that is determined by the total 
and the local entropies. In this sense, the correlation between the visible and the hidden layer 
may become another good indicator to the RBM performance.

We also investigated the work done on the RBM by switching the parameters of the energy function.  
The trajectories of the visible and hidden vectors in configuration space were generated using 
Markov chain Monte-Carlo simulation. The distribution of the work follows the Gaussian distribution and 
its standard deviation grows with training epochs. We discussed the Jarzynski equality, which 
connects the free energy difference and the average of the exponential function of the work 
over the trajectories.

A more detailed analysis from a full thermodynamics or statistical physics point of view can bring 
us useful insights into the performance of RBM. This course of study may enable us to come up with 
possible methods for a better performance of RBM for many different applications in the long run. 
Therefore, it may be worthwhile to further pursue our study, e.g. a rigorous assessment of scaling 
behavior of thermodynamic quantities with respect to epochs as the sizes of the visible and hidden
layers increase. We also expect that a similar analysis on a quantum Boltzmann machine can be 
valuable as well.

%\appendix
%\section{}

%\begin{acknowledgments}
%\end{acknowledgments}

\bibliography{rbm_manuscript}
\end{document}